\documentclass[prl,twocolumn]{revtex4}
\usepackage{graphicx}
\usepackage{amsmath}
\usepackage{amsfonts}
\usepackage{amssymb}%
\begin{document}
\title{Revisiting the Mn-doped Ge\\ 
 using the Heyd-Scuseria-Ernzerhof hybrid functional}
\author{A. Stroppa$^{1}$, G. Kresse$^2$ and  A. Continenza$^{3}$}

\affiliation{$^1$ CNR-SPIN
L'Aquila, Italy} 
\affiliation{$^2$ Faculty of Physics,  and Center for
Computational Materials Science, Universit\"at Wien, Sensengasse
8/12, A-1090 Wien, Austria}
\affiliation{$^3$ CNISM- Dipartimento di Fisica Universit\`a degli Studi di
L'Aquila,\\ Via Vetoio 10 L'Aquila, Italy}
\begin{abstract}

We perform a comparative \textit{ab-initio} study of Mn-doped  Germanium semiconductor 
using the  Perdew-Burke-Ernzerhof (PBE) exchange-correlation functional, DFT+$U$ and
Heyd-Scuseria-Ernzerhof hybrid functional
(HSE). We show that the HSE functional is able to correctly account 
for  the relevant  ground state properties of the host matrix  as well
 as of  Mn-doped semiconductor. Although the  DFT+$U$ and the HSE 
 description are very similar, some  differences still remain.
In particular, the half-metallicity is lost using DFT+$U$ when a suitable 
$U$ value,  
tuned to recover the photoemission spectra, is employed. 
For comparison, we also discuss the case of Mn in Silicon. 
\end{abstract}
\pacs{}

\maketitle

\section{Introduction} 
   Dilute magnetic semiconductors (DMSs) are  still a
topic of great current interest.\cite{Jungwirth,Zutic,ohno1,awschalom} 
The theoretical description of the interaction of transition metal doped 
semiconductors is  challenging since  localized states interact
significantly with delocalized states. Density functional theory (DFT) 
in the local density or generalized gradient  approximation (LDA or GGA) 
for the exchange-correlation energy is not able to properly describe the 
non-locality of the screened exchange interaction and, furthermore, possesses a sizeable 
self-interaction error.\cite{kummel} These limitations are particularly severe in the
case of  localized orbitals, \textit{e.g.} Mn-$3d$ states, which are described as too
shallow in energy resulting in a large 
hybridization with anion $p$-states. As a result, the Mn-$3d$ states are  over-delocalized.
The situation is particularly serious in the case of small band-gap semiconductors
(such as Ge) which are described as 
metals in  LDA/GGA, thus producing an overestimated hybridization among the valence 
and conduction Ge states and  Mn-$d$ states.
The   physics of localized $d$ states can be partially described  using a  
DFT+$U$ formalism, which introduces a local correction $U$ to recover the proper
position of the Mn-$d$ states\cite{LOT}. However, in 
Ge, the accurate electronic 
properties are not completely recovered: \textit{e.g.} the half-metallicity of the compound is lost within DFT+$U$ scheme.

Very recently, hybrid Hartree-Fock density functionals, 
which mix a fraction of the exact Fock exchange with the DFT exchange, 
have been widely applied to extended solid state systems.\cite{kummel,GaN,StroppaTMO,Togo,Kresse1,Kresse2,Scan1,Scan2,Scan3,Scan4,ZungerLast1,ZungerLast2,VanDeWalleLast,AleMF,ZungerLast,Cesare1,Cesare2,Hummer}
In this paper, we mainly focus on 
 Mn-doping in bulk Ge by performing  hybrid-density functional theory calculations.
We will show that the HSE functional gives a satisfactory description of the  structural, electronic and magnetic properties of Ge-based DMS, consistent with  
experimental data. Furthermore, for few selected properties, and, for sake of comparison, we will also include some results of Mn-doped Silicon.  

\section{Computational details}

The calculations were performed within the projector augmented-wave
(PAW) method\cite{paw} using the Perdew-Burke-Ernzerhof (PBE) generalized gradient approximation (GGA)\cite{pbe} and
Heyd-Scuseria-Ernzerhof (HSE) hybrid
functional\cite{computational,erratum}, recently implemented in the VASP
code.\cite{computationalvasp,krakau}  We also used the DFT+$U$ method for Mn within Dudarev's approach\cite{Dudarev}
fixing $U$ to 6 eV and $J$ to 1 eV (PBE was used for the DFT part).
The kinetic energy cutoff used for the orbitals was set to  300 eV.
Monkhorst-Pack $k$-point grids of $10\times 10 \times 10$ and $6 \times 6 \times 6$
were used 
to sample the Brillouin zone  of the Ge-bulk and of the 64-atom
unit cell, respectively. All the atomic internal
positions were relaxed.  
In the following, we will focus on the Ge bulk system, single 
(substitutional and interstitial) and double substitutional (dimer)
 Mn impurities in a 64-atom Germanium cell. For Silicon, we will consider the bulk case and the Mn substitutional impurity in a 64-atom unit cell.

\section{Bulk Ge}
For bulk Ge, the  calculated equilibrium properties 
are in good  agreement with experiments:
the HSE lattice constant is 
5.703 (5.792) \AA\ using HSE (PBE)
within 0.7 (2.3) \% of the experimental value of 5.660 \AA;\cite{Scheffler}
the HSE bulk modulus (731 kbar) improves over the PBE value (571 kbar) when compared to
experiment (768 kbar\cite{landolt}). Furthermore, we remark that
within HSE the energy gap is properly described to be indirect 
(0.63 eV including SOC\cite{peralta} 
compared to the experimental value of  0.74 eV\cite{madelung}).
A similar result was obtained within HSE for Si\cite{peralta}:
the HSE lattice constant is 5.444 \AA \ using HSE\cite{peralta} compared to the
experimental one of 5.430 \AA.\cite{landolt} The calculated indirect energy gap is 1.12 eV\cite{peralta}  while the experimental one is 1.17 eV.\cite{landolt}
It is interesting to note that, for Silicon,  self-interaction schemes
 which are  often used to improve
the electronic structure description,\cite{sic,kummel}
do not open the gap.\cite{filippetti}
This is important for our present study,
since a faithful description of the equilibrium  
properties of the host semiconductor is at the basis  of an appropriate
 description of the doped system.

\section{Mn impurities in Ge} 
In Tab.\ref{MnGe}, we summarize our main results, {\em i.e.} 
 formation energy\cite{VanDeWalle} ($\Delta\textrm{H}$),
 Mn-Ge bond length
($d_{\textrm{Mn-Ge}}$) and  the Mn magnetic moment ($\mu$)  at their respective theoretical lattice
constants for the considered Mn-doping cases. The formation energies are evaluated
with respect to the calculated Ge and Mn equilibrium bulk phases (diamond Ge and AFM-fcc Mn).
Of course, while Ge-rich growth conditions can be safely assumed to fix the
Ge chemical potential to its bulk value, the same is not true for Mn so that
the formation energy is a function of the Mn chemical potential
$\mu_{\textrm{Mn}}$. In  Table~\ref{MnGe} we report the value for Mn-rich
conditions fixing $\mu_{\textrm{Mn}}$ to the corresponding bulk value ($\alpha$-Mn).
%

 We note that the experimental
evidence of  \textit{local} Ge-lattice dilation upon Mn-doping is correctly described
using HSE yielding a Mn-Ge distance 2 \% larger than
the ideal Ge-Ge bond-length,
while PBE gives a local contraction of $-$2\%\, 
 and DFT+$U$ finds a smaller local dilation of 
0.4 (0.8)\%\ when using the theoretical PBE (HSE or experimental)
 lattice constant.\cite{kettaps,Tsui1,Tsui} The most recent extended x-ray absorption fine
structure 
results\cite{LOT1,LOT2,Pochet} yield a Mn-Ge coordination distance of
2.50-2.51$\pm$ 0.03 \AA \ for the samples obtained at low temperature
and which are thought to be  best candidates for Mn occupation on substitutional sites\cite{LOT2}. Clearly, these results match  the HSE result and 
 the DFT+$U$ as well.  Similar results for bond-lengths contraction/dilation within 
different DFT schemes were reported also for III-V based DMS\cite{Furdyna}.
%

\begin{figure}
\includegraphics[scale=.5,angle=0]{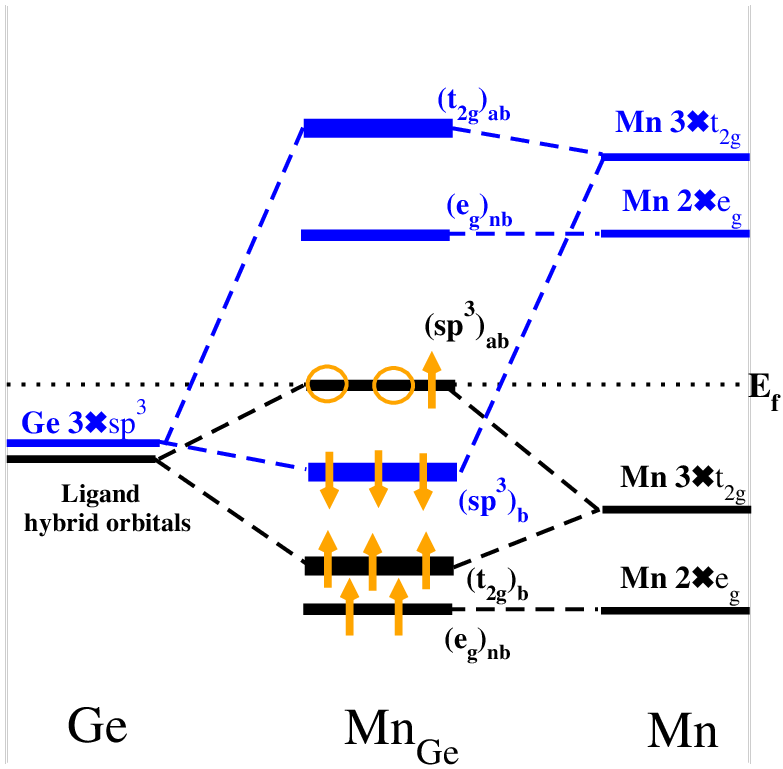}
\vspace{1cm.}
\caption{(Color on line) Molecular energy diagram of the Mn-$d$ states (right) 
interacting with the Ge-$sp^{3}$ hybrid orbitals (left). 
The Mn-$s$ states are not shown for clarity (see text). 
$b$, $ab$ and $nb$ subscripts label bonding, anti-bonding and non-bonding orbitals. 
Arrows denote up/down electrons 
while circles indicate holes. The central panel (Mn$_{\textrm{Ge}}$)
shows $p-d$ hybridization for substitutional Mn at Ge site.}
\label{orbit}
\end{figure}

A simple molecular orbital description, as sketched in Fig.~\ref{orbit}, can  be useful in order
to describe the interaction of Mn in the tetrahedral Ge ligand field, as also done previously
 for similar compounds\cite{Azunger}.  
We recall that in diamond like semiconductors,
 the $sp$ valence states arrange to form $sp^{3}$ hybrid orbitals, 
 each of them filled with a bonding electron pair.
 If one Ge atom is removed creating a Ge vacancy, 
 4 $sp^{3}$ hybrids point towards the vacant Ge atom,
   each filled with one electron (dangling bonds). 
The Ge vacancy is now replaced by a Mn atom.
 Due to the local tetrahedral symmetry, the Mn $d$-states 
are split into 3-fold degenerate $t_{2g}$ and 2-fold degenerate
$e_{g}$-like states, further splitted by the local exchange field (see  Fig.~\ref{orbit}, right part).   From linear combinations of the four $sp^3$
Ge dangling bonds pointing towards the
transition-metal impurity, an $s$-like $a_{1\uparrow,\downarrow}$ orbital 
and three $p$-like $t_{2g\uparrow,\downarrow}$  orbitals are
formed.   The
host  $a_{1\uparrow,\downarrow}$ orbital and the transition-metal $4s_{\uparrow,\downarrow}$ states form a doubly 
occupied bonding state deep in the
semiconductor valence band, and an empty antibonding state high in the conduction band (not included in Fig.~\ref{orbit}).  
For the majority component,  the  Mn-$t_{2g}$ orbitals  are lower in energy than the Ge-$t_{2g}$ $sp^{3}$ hybrid states. They interact  giving rise to
3 bonding states (3$\times$Mn-$t_{2g}$)$_{b}$ (see Fig.~\ref{orbit})  and 3 antibonding states
(3$\times$Ge-$sp^{3}$)$_{ab}$. The Mn-$e_{g}$ states do not hybridize because they are non-bonding in 
a tetrahedral ligand field.  For the minority component, the Ge-$t_{2g}$ $sp^{3}$ orbitals
are lower in energy than the Mn-$t_{2g}$ states. Upon interaction, 
they give rise to 3 bonding orbitals (3$\times$Ge-$sp^{3}$)$_{b}$
and 3 anti-bonding orbitals (3$\times$Mn-$t_{2g}$)$_{ab}$. 
The complex is characterized by a total of 
11 electrons (4 from the nearest Ge atoms and 7 from the Mn impurity atom). 
Disregarding the two electrons occupying the
lowest $a_1$ symmetry-like state, one needs to fill the orbitals with nine electrons
as shown in Fig.~\ref{orbit}:  Clearly the Mn impurity is in a high spin state with 5 
$d$-electrons 
in the majority channel
(2 electrons in the $e_{g}$ and 3 in the $t_{2g}$-like states) 
and zero $d$-electrons in the minority channels.   However, while the minority
valence Ge-$sp^{3}$  states are fully occupied, the majority
states accommodate two holes.
This simple molecular picture suggests that: 
i) the compound is half-metallic,
with the Fermi level falling within the Ge majority valence band; 
ii) the total spin moment  of the complex, \textit{i.e} $n_{\uparrow}-n_{\downarrow}$,
is 3 $\mu_{B}$; iii) 
the local Mn $d$ spin moment is  5 $\mu_{B}$ partially compensated by 
the holes in the $sp^{3}$ states;
iv) the induced spin 
moment on the 4 nearest  Ge atoms, \textit{i.e.}  
$n^{{\rm Ge}-sp^{3}}_{\uparrow}-n^{{\rm Ge}-sp^{3}}_{\downarrow}$ should be sizeable
and opposite to the spin on the Mn atom.
In line with previous calculations~\cite{schult,aless,park}, the calculated results (see Tab.~\ref{MnGe}) confirm this picture
finding a total spin moment of exactly 3 $\mu_{B}$ in the unit cell, 
4.1 $\mu_{B}$ at the Mn atom, and 
 $-$0.11\ $\mu_{B}$ at the nearest Ge-atoms. 
Furthermore the sizeable induced moments on Ge atoms  suggest that the holes are rather  delocalized.

\begin{figure}
\includegraphics[scale=0.5,angle=0]{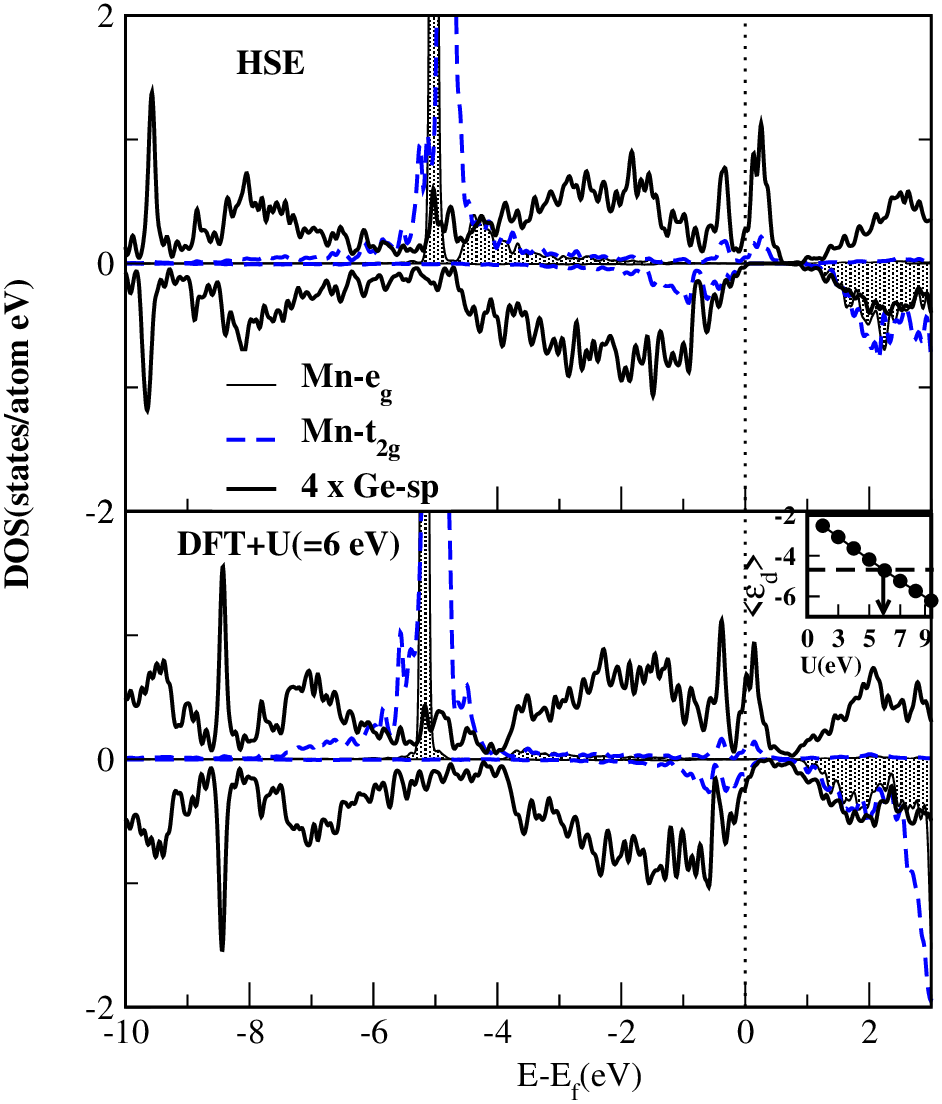}
\vspace{1cm.}
\caption{Density of states projected on the Mn impurity site (top panel)
 in symmetry resolved angular momentum components: $t_{2g}$ (dashed line) and $e_g$ 
 (shadow) states
for Mn substitutional impurity.
The density of states projected on the $l=1$ component of the 4
Ge (bottom panel) coordinated with the Mn impurity is also shown (solid line). 
The inset shows the Mn-$d$ center of mass as a function of the $U$ value. The horizontal line
indicates the value found within HSE.}
\label{DOS}
\end{figure}

The local angular momentum decomposed density of states (DOS) shown in Fig.~\ref{DOS} 
is  consistent with this orbital interaction diagram.
The top panel shows the HSE results, whereas the bottom panel 
reports  DFT+$U$ results for $U$=6 eV.
In the inset, we show the relation between 
the center of mass of the Mn-$d$ majority states, $\langle \epsilon_d \rangle$, 
and the $U$ value.
The horizontal line indicates the $\langle  \epsilon_d \rangle$ value, which matches
 the HSE result ($U$=6 eV).  
Fig.~\ref{DOS} clearly confirms the interpretation discussed above.
 In particular,  integration of the majority
total density of states from the Fermi level up to the end of the Ge-valence
band exactly sums up to  2 electrons: these are the two holes
required to fill the Ge valence band. 
Mn-substitution into the Ge host matrix  does not produce
a Jahn-Teller ion, 
but rather a  Mn$^{2+}$ ionic state with a $d^{5}$ configuration and 
two spin-polarized holes.
Let us now compare the HSE and DFT+$U$ DOS.
Due to the choice of the $U$ value, the Mn-$d$ states have the same energy
within HSE and DFT+$U$
but the hybridization between Ge-$sp$ and Mn $t_{2g}$ and $e_{g}$ states (this
latter symmetry-allowed away from $\Gamma$) is underestimated within DFT+$U$
compared to HSE. While the on-site $U$ mainly localizes the Mn $d$
states, HSE also acts through the screened exchange 
on Ge-$p$ states, lowering their energy position 
and leading to a larger Mn-$d$ Ge-$p$ hybridization.
As a matter of fact, the larger hybridization in HSE compared to DFT+$U$
can be recognized 
just below $-4$~eV, where  a peak
in the $e_g$ character (shaded region in Fig.~\ref{DOS})  is completely
absent in the present DFT+$U$ description. 
We note that previous LDA+$U$ calculations\cite{LOT} with $U$=4 eV, although
reproducing the peak  at $-4$~eV characteristic of the Mn-Ge bond,\cite{LOT}
 gave a quite different
density of states for both the $t_{2g}$ and $e_{g}$ states, which is due to 
 the strong dependence   of the localized $d$-states
description on the $U$ parameter.
Finally, as found in Ref.\ \onlinecite{LOT},  
the half-metallic character of the compound is destroyed within
 DFT+$U$:  the  energy position of the Mn-$t_{2g}$ Ge-$sp$ 
 bonding minority states,
 whose energy position is mainly determined by the atomic Ge-$sp$ levels,   
is raised towards higher energies causing an incomplete filling of the 
minority valence band.

\begin{figure}
\includegraphics[scale=.5,angle=0]{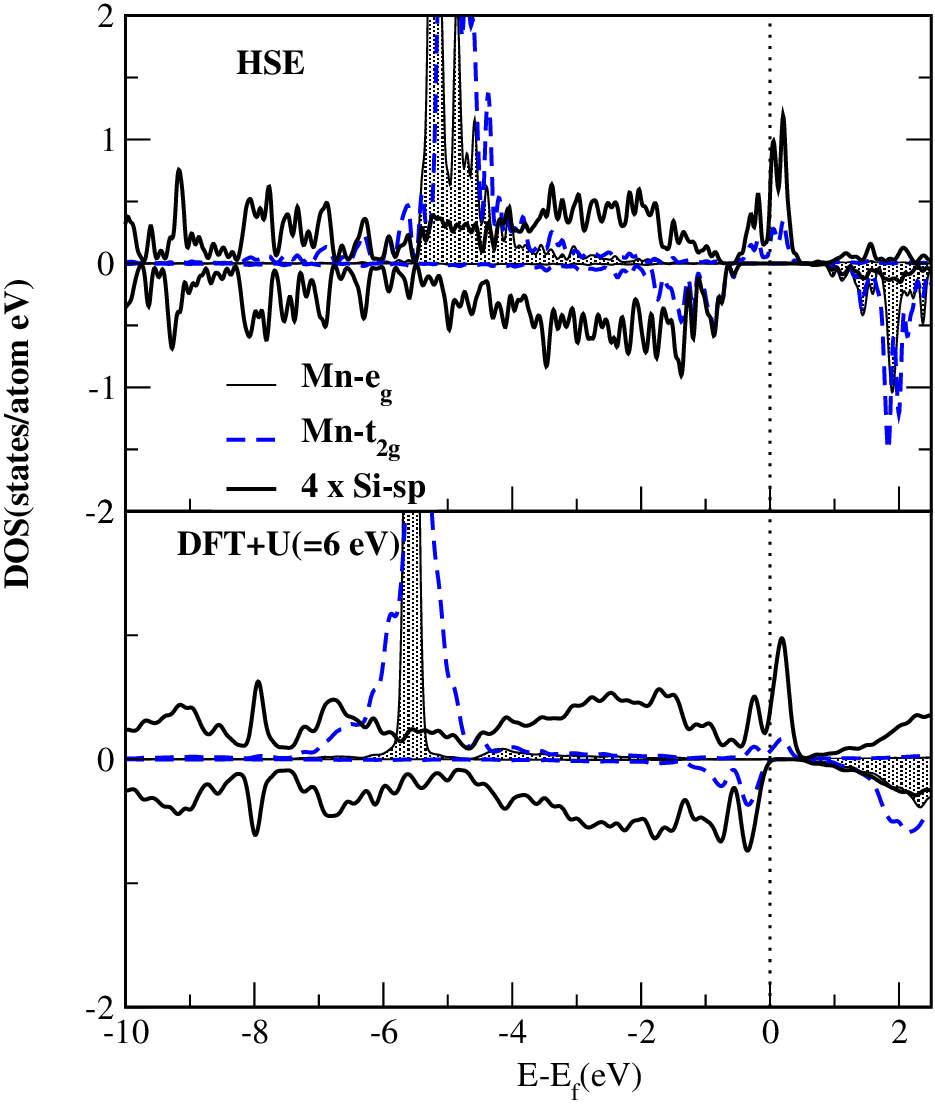}
\vspace{1cm.}
\caption{(Color on line)  HSE Density of states projected on the Mn impurity 
site (top panel) for Mn in Si. Labels as in Fig.\ \ref{DOS}. DFT+$U$, $U$=6 eV  (bottom panel).}
\label{dos_Si}
\end{figure}

In a previous study,\cite{aless}  it was shown that the half-metallicity is favored 
in Mn doped Ge while in Silicon matrix it is lost.  
In Fig.~\ref{dos_Si}, we show the HSE DOS for substitutional Mn$_{\textrm{Si}}$
(top panel) and DFT+$U$, with $U$=6 eV, the same used for Mn$_{\textrm{Ge}}$
(bottom panel). The correction for the 
 self-interaction error has a larger effect on Si-$sp^{3}$ states, since they are quite localized. Therefore, they are pushed down in energy. According to the  orbital energy diagram shown in Fig.\ \ref{orbit}, also the minority \textit{bonding}
 (Si-$sp^{3}$)$_{b}$ are shifted down in energy, favoring the half-metallicity.
Obviously, the Mn-$d$ states are also corrected for the self-interaction error, and they are 
pushed down in energy as well
On the other hand, DFT+$U$ corrects only the Mn-$d$ states, but not the Si-$sp^{3}$ states.
This gives a near-half metallic structure and an underestimation of the hybridization of 
Mn-Si states compared to the HSE description.

\begin{table} 
\caption{Formation energy $\Delta \textrm{H}$, Mn-X distances $d_{\textrm{Mn-X}}$ and
magnetic moments $\mu$ for Mn-doped Ge for various structures. Distances in
 parentheses specify the ideal Ge-Ge bond-length of the host (lines $d_{\textrm{Mn-X}}$). 
 Local Mn magnetic moment as well as total magnetic moment (in parentheses) are specified.}
\vspace{0.5cm}
\begin{tabular}{cccc} \hline \hline 
                                & PBE           & DFT+$U$       &   HSE           \\\hline \hline
{\bf Mn substitutional site}    &               &                &                 \\  
$\Delta \textrm{H}$ (eV/Mn)              &  1.5          &                &   0.9           \\
$d_{\textrm{Mn-Ge}}$ (\AA)               &  2.46 (2.51)  &  2.53 (2.51)   &   2.52 (2.47)   \\ 
 $\mu$ ($\mu_B$)                &  3.3  (3.1)   &  4.1  (3.4)    &   4.1  (3.0)    \\
\hline
{\bf Mn interstitial site}         &               &                &                 \\  
$\Delta \textrm{H}$ (eV/Mn)              &  2.1          &                &  1.8            \\
$d_{\textrm{Mn-Ge}}$ (\AA)               &  2.57 (2.51)  &  2.63 (2.51)   &  2.58 (2.47)    \\
$\mu$ ($\mu_B$)                 &  3.4 (4.0)    &   4.2 (4.8)    &   3.8 (4.1)     \\
\hline
{\bf Mn-Mn dimer}               &               &                &                 \\  
$\Delta \textrm{H}_{\textrm{AFM}}$ (eV/Mn-pair)   &  2.9          &                &   1.5           \\
$\Delta \textrm{H}_{\textrm{FM}}-\Delta \textrm{H}_{\textrm{AFM}}$ &  0.77         &   0.22         &   0.15          \\
$d_{\textrm{Mn-Mn (FM)}} $ (\AA)          &  2.55 (2.51)  &   2.97 (2.47)  &   2.99  (2.47)  \\
$d_{\textrm{Mn-Mn(AFM)}}$ (\AA)         &  1.95 (2.51)  &   2.71 (2.47)  &   2.83  (2.47) \\
\hline\hline
\end{tabular}
\label{MnGe}
\end{table}

\subsection{The single interstitial impurity}  This defect in a Germanium matrix
possesses twice the formation energy
as the substitutional impurity (HSE), hence
it is unlikely to form.  Here we only note that a tendency to a local expansion 
around Mn is found for all functionals.  Interestingly, 
the local magnetic moment is larger for DFT+$U$  than for HSE, 
suggesting sizeable differences in the interaction of the impurity  with the local environment.

\subsection{Mn-dimers} Double Mn 
substitutions on two nearest-neighbouring Ge sites are of
particular interest, since they are inferred to occur  at
experimental growth conditions~\cite{park,LOT1} leading to 
nucleation of Mn-precipitates. In addition, they might be also detrimental
for the magnetic ordering, since dimers show antiferromagnetic (AFM) coupling
with no net spin moment.
Unfortunately, GGA-based
calculations are not entirely conclusive, since the dimer configuration
becomes stable at a too  small bond-length (1.95 \AA) not compatible with 
the Mn ionic radius  ($\simeq$ 1.1-1.3 \AA) or bond distances in Mn-Ge compounds.
Thus,  published results~\cite{park,imp2} often refer to the ideal
unrelaxed structure which, of course, strongly overestimates the heat of 
formation of the dimer.
From the results reported in Tab.~\ref{MnGe}, assuming 
thermodynamic equilibrium,
we can comment on the relative concentration of single substitutional sites and dimers.
 At thermodynamic equilibrium, the concentration $c$ of a defect with $n_{\rm Mn}$ Mn atoms is roughly proportional to $e^{-(\Delta \textrm{H}- n_{\rm Mn}\Delta \mu_{\rm Mn})/k_{B}T}$ where $\Delta$H is the formation energy, $k_{\textrm{B}}$ the Boltzman constant, and $T$ the temperature. Supposing  $\Delta \mu_{\textrm{Mn}}\approx0$~eV (thermodynamic equilibrium with $\alpha-$Mn), the probability of finding a substitutional Mn or dimer is $e^{-0.9/k_{B}T}$ and $e^{-1.5/k_{B}T}$, respectively, 
\textit{i.e.} monomers are more likely to form than dimers. More generally, for a specific $\Delta \mu_{\textrm{Mn}}$ 
the probabilities for single substitutions and dimers are,
\[
e^{(-0.9+\Delta \mu_{\textrm{Mn}})/k_{B}T} \quad \mbox{and} \quad e^{(-1.5+2\mu_{\Delta \textrm{Mn}})/k_{B}T}.
\]
Therefore the dimer concentration is larger than the monomer concentration only for 
$\Delta \mu_{\textrm{Mn}}>0.6$~eV, \textit{i.e.} at extremely Mn-rich conditions, 
where $\alpha-$Mn precipitates are anyway already preferred over the formation of monomers (or dimers).
Although kinetic effects might well hinder the nucleation of larger precipitates, 
our calculated thermodynamics suggests a rather low dimer concentration.
It is important to note, that the thermodynamic arguments alone, presented here,
are not enough to fully discuss the relative probability of occurrence 
of the monomers with respects to dimers, as these systems are usually 
grown out of the thermodynamic equilibrium and kinetic effects may have an important role.


Finally, the stabilization energy of  AFM over  FM coupling of 
nearby impurities is much lower within HSE than GGA.
Furthermore, for HSE (and DFT+$U$),
the calculated Mn-Mn distance for  both FM and
AFM magnetic alignment, are in line with experiments ---always reporting local
lattice dilation --- as well as in line with the 
Mn-Mn distances  in the FM Mn$_5$Ge$_3$ compound (varying between 2.52 
and 3.06 \AA)~\cite{mn5ge3,mn5ge3.stroppa.1,mn5ge3.stroppa.2}.
 
\section{Conclusions} In summary,  
we have performed a comparative study of   substitutional Mn$_{\textrm{Ge}}$
by using PBE, PBE+$U$ and HSE functional. The main focus is on the differences arising 
from three different treatments of the exchange-correlation term, namely the PBE, DFT+$U$,
 and HSE. As  well known, the PBE treatment can not describe  satisfactorily 
the ground state properties of the Mn in the host semiconductor matrix. Including the $U$ correction at DFT level improves the description. However, some differences still remain when compared to HSE.  For example, 
the HSE  Mn-$d$ peak position is found at $\sim-$5 eV with respect to the Fermi energy, \textit{i.e.} in the same 
energy region as observed in  photoemission experiments.\cite{LOT}  When using  the DFT+$U$ method and fixing the $U$ parameter in order to recover the experimental  $d$-peak position, the hybridization with Ge-$p$ states around $-$4 eV  is underestimated compared to HSE and the experimental photo-emission 
peak.\cite{LOT}  Furthermore for the same $U$, 
the half-metallic character is not predicted by DFT+$U$.
The fact that HSE accurately describes the host semiconductor and, at the same time,  the interaction of localized Mn states with the host valence states
makes this functional a valuable approach for studying transition metal defects in semiconductors. It is also true that the HSE calculations are usually quite more 
computationally demanding with respect to DFT+$U$. Therefore, whenever  
a compromise between accuracy and computation effort is required in the calculations, 
a preliminary HSE study may be useful for choosing an appropriate $U$ value, which is often not accessible from experiments.

%

\acknowledgments
This work was supported by the Austrian {\em Fonds
zur F\"orderung der wissenschaftlichen Forschung} and by a computing
grant at CINECA-HPC center.

\bibliography{biblio}

\end{document}